\documentstyle[12pt,epsf]{article}

\catcode`\@=11
\long\def\@makefntext#1{
\protect\noindent \hbox to 3.2pt {\hskip-.9pt
$^{{\ninerm\@thefnmark}}$\hfil}#1\hfill}                

 \def\@makefnmark{\hbox to 0pt{$^{\@thefnmark}$\hss}}  

\def\ps@myheadings{\let\@mkboth\@gobbletwo
\def\@oddhead{\hbox{}
\rightmark\hfil\ninerm\thepage}
\def\@oddfoot{}\def\@evenhead{\ninerm\thepage\hfil
\leftmark\hbox{}}\def\@evenfoot{}
\def\sectionmark##1{}\def\subsectionmark##1{}}

\newcounter{sectionc}\newcounter{subsectionc}\newcounter{subsubsectionc}
\renewcommand{\section}[1] {\vspace{0.6cm}\addtocounter{sectionc}{1}
\setcounter{subsectionc}{0}\setcounter{subsubsectionc}{0}\noindent
	{\bf\thesectionc. #1}\par\vspace{0.4cm}}
\renewcommand{\subsection}[1] {\vspace{0.6cm}\addtocounter{subsectionc}{1}
	\setcounter{subsubsectionc}{0}\noindent
	{\it\thesectionc.\thesubsectionc. #1}\par\vspace{0.4cm}}
\renewcommand{\subsubsection}[1]
{\vspace{0.6cm}\addtocounter{subsubsectionc}{1}
	\noindent {\rm\thesectionc.\thesubsectionc.\thesubsubsectionc.
	#1}\par\vspace{0.4cm}}

\newcounter{appendixc}
\newcounter{subappendixc}[appendixc]
\newcounter{subsubappendixc}[subappendixc]

\renewcommand{\appendix}[1] {\vspace{0.6cm}
	\refstepcounter{appendixc}
	\setcounter{figure}{0}
	\setcounter{table}{0}
	\setcounter{equation}{0}
	\renewcommand{\thefigure}{\Alph{appendixc}.\arabic{figure}}
	\renewcommand{\thetable}{\Alph{appendixc}.\arabic{table}}
	\renewcommand{\theappendixc}{\Alph{appendixc}}
	\renewcommand{\theequation}{\Alph{appendixc}.\arabic{equation}}
	\noindent{\bf Appendix \theappendixc #1}\par\vspace{0.4cm}}

\def\abstracts#1{{
	\centering{\begin{minipage}{30pc}\tenrm\baselineskip=12pt\noindent
	\centerline{\tenrm ABSTRACT}\vspace{0.3cm}
	\parindent=0pt #1
	\end{minipage}}\par}}

\renewenvironment{thebibliography}[1]
	{\begin{list}{\arabic{enumi}.}
	{\usecounter{enumi}\setlength{\parsep}{0pt}
\setlength{\leftmargin 1.25cm}{\rightmargin 0pt}
	 \setlength{\itemsep}{0pt} \settowidth
	{\labelwidth}{#1.}\sloppy}}{\end{list}}

\newcounter{itemlistc}
\newcounter{romanlistc}
\newcounter{alphlistc}
\newcounter{arabiclistc}

\newcommand{\fcaption}[1]{
	\refstepcounter{figure}
	\setbox\@tempboxa = \hbox{\tenrm Fig.~\thefigure. #1}
	\ifdim \wd\@tempboxa > 6in
	   {\begin{center}
	\parbox{6in}{\tenrm\baselineskip=12pt Fig.~\thefigure. #1}
	    \end{center}}
	\else
	     {\begin{center}
	     {\tenrm Fig.~\thefigure. #1}
	      \end{center}}
	\fi}

\newcommand{\tcaption}[1]{
	\refstepcounter{table}
	\setbox\@tempboxa = \hbox{\tenrm Table~\thetable. #1}
	\ifdim \wd\@tempboxa > 6in
	   {\begin{center}
	\parbox{6in}{\tenrm\baselineskip=12pt Table~\thetable. #1}
	    \end{center}}
	\else
	     {\begin{center}
	     {\tenrm Table~\thetable. #1}
	      \end{center}}
	\fi}

\def\@citex[#1]#2{\if@filesw\immediate\write\@auxout
	{\string\citation{#2}}\fi
\def\@citea{}\@cite{\@for\@citeb:=#2\do
	{\@citea\def\@citea{,}\@ifundefined
	{b@\@citeb}{{\bf ?}\@warning
	{Citation `\@citeb' on page \thepage \space undefined}}
	{\csname b@\@citeb\endcsname}}}{#1}}

\newif\if@cghi
\def\cite{\@cghitrue\@ifnextchar [{\@tempswatrue
	\@citex}{\@tempswafalse\@citex[]}}
\def\citelow{\@cghifalse\@ifnextchar [{\@tempswatrue
	\@citex}{\@tempswafalse\@citex[]}}
\def\@cite#1#2{{$\null^{#1}$\if@tempswa\typeout
	{IJCGA warning: optional citation argument
	ignored: `#2'} \fi}}

\def\fnt#1#2{\footnotetext{\kern-.3em
	{$^{\mbox{\sevenrm #1}}$}{#2}}}

 1
 1
 1

\font\tenbf=cmbx10
\font\tenrm=cmr10
\font\tenit=cmti10

\font\ninerm=cmr9

\begin{titlepage}\begin{flushright}
NORDITA - 94/65 N,P
\end{flushright}
\begin{center}
\vspace{3cm}
{\large\bf EFFECTIVE LAGRANGIANS
FOR LIGHT QUARKS$^*$}\\[2cm]
Johan Bijnens\\[1cm]
NORDITA, Blegdamsvej 17\\
DK 2100 Copenhagen \o, Denmark
\end{center}
\vspace{3cm}
\begin{abstract}
The class of phenomenological Lagrangians used for light
constituent quarks is discussed.
The Nambu-Jona-Lasinio model is then argued to be a good phenomenological
choice and the quality of its prediction in the purely hadronic sector
including several relations
between parameters illustrated.
Then we use these models to calculate nonleptonic matrix elements.
The $\pi^+-\pi^0$ mass difference and the $B_K$ parameter, relevant in
$K^0-\overline{K}^0$ mixing, are treated in more detail.
\end{abstract}
\vspace{3cm}
November 1994\\
$^*$ {Invited talk at Perspectives in
Particle Physics 94, Brijuni, Croatia, Sept. 1994}
\end{titlepage}

\begin{document}
\newcommand{\rcite}[1]{{\cite{#1}}}
\newcommand{\rref}[1]{{(\ref{#1})}}
\newcommand{\tref}[1]{{\ref{#1}}}
\newcommand{\rlabel}[1]{{\label{#1}}}
\newcommand{\rbibitem}[1]{\bibitem{#1}}
\newcommand{\be}{\begin{equation}}
\newcommand{\ee}{\end{equation}}
\newcommand{\ba}{\begin{eqnarray}}
\newcommand{\ea}{\end{eqnarray}}
\newcommand{\dis}{\displaystyle}
\newcommand{\ovpi}{{\overline{\Pi}}}
\newcommand{\qvev}{\langle\overline{q}q\rangle}
\newcommand{\Pids}{\Pi_{\Delta S =2}}

\centerline{\tenbf EFFECTIVE LAGRANGIANS}
\baselineskip=22pt
\centerline{\tenbf FOR LIGHT QUARKS}
\vspace{0.8cm}
\centerline{\tenrm JOHAN BIJNENS}
\baselineskip=13pt
\centerline{\tenit NORDITA, Blegdamsvej 17}
\baselineskip=12pt
\centerline{\tenit DK 2100 Copenhagen \o, Denmark}
\vspace{0.9cm}
\abstracts{The class of phenomenological Lagrangians used for light
constituent quarks is discussed.
The Nambu-Jona-Lasinio model is then argued to be a good phenomenological
choice and the quality of its prediction in the purely hadronic sector
including several relations
between parameters illustrated.
Then we use these models to calculate nonleptonic matrix elements.
The $\pi^+-\pi^0$ mass difference and the $B_K$ parameter, relevant in
$K^0-\overline{K}^0$ mixing, are treated in more detail.}
\vfil
\rm\baselineskip=14pt
\section{Introduction}
The problems of dealing with the strong interaction at low and intermediate
energies are well known. At short distance we can use perturbative Quantum
Chromo Dynamics (QCD) but due to asymptotic freedom this can no longer be
done at low energies. The coupling constant there becomes too large.
A general method, that is however extremely manpower and computer intensive,
is using lattice gauge theory methods. An overview of this field can be found
elsewhere in these proceedings\rcite{andreas}.

At very low energies we can use the methods of Chiral Perturbation
Theory (CHPT).
A good overview of the present state of the art here can be found in the
DA$\Phi$NE workshop report\rcite{CHPT}.
CHPT is a rigorous consequence of the symmetry pattern
in QCD and its spontaneous breaking. Both perturbative QCD and CHPT are good
theories in the sense that it is in principle possible to go to higher orders
and calculate unambiguously. The size of the higher orders also gives an
estimate of the expected accuracy of the result. A disadvantage of CHPT is
that as soon as we start going beyond lowest order the number of free
parameters increases very rapidly thus making calculations beyond the lowest
few orders rather impractical. We would thus like to obtain these free
parameters directly from QCD.

This has so far been rather difficult to do. The reason is that all
available approaches,like lattice QCD, QCD sum rules, etc.\ ,
have problems with enforcing the correct chiral behaviour. We would also
like to understand the physics behind the numbers from the lattice
calculations in a more intuitive fashion. Therefore there is a need for
some models that interpolate between QCD and CHPT. We will require that these
models have the correct chiral symmetry behaviour.

It should be kept in mind that these are models and not QCD. The hope is
that these models will catch enough of the essential part of the behaviour
of QCD at low energies that they can be useful.
Two major classes exist, those with
higher resonances than the pseudoscalars included and staying at the
hadronic level, or those with some kind of quarks.  Both of these have their
drawbacks. In the first case there still tends to be a large number of
parameters and in the second case most models do not include
confinement.
Confinement
is treated by explicitly looking only at colour singlet
observables. The other drawback is inherent in the use of a model. It is not
possible to systematically expand and get closer to the ``true'' answer.

We will here look at models including some kind of constituent quarks. The
main
motivation here is that the standard constituent quark picture
explains the hadron spectrum rather well. It has problems
when interactions have to be included. It also tends to
break chiral symmetry explicitly. Here we do not attempt to explain
the hadron spectrum but instead focus on the few lowest lying states only.

The class of models we will look at, is those where the fundamental
Lagrangian contains quarks and sometimes also explicitly meson fields.
There exists a whole set of these models of increasing sophistication.
Models that are mainly for study of the spectrum like the bag model are not
included. See \rcite{Donoghue} for a review of various aspects
of this whole area.

The lowest member of the hierarchy are the quark-loop models. Here the basic
premise is that interactions of mesons proceed only via quark loops. The
kinetic term for the mesons is added by hand. As a rule these models have
some problems with chiral symmetry. In particular pointlike couplings
of more than one meson to a quark-antiquark pair have to be added in order
to be consistent. This goes under various names like bare-quark-loop model.
A version that incorporates chiral symmetry correctly and also considers
gluons is known as the Georgi-Manohar model\rcite{GM}. Another variation
is to use the linear sigma model coupled to quarks.

The next level is what I would call improved quark-loop models. Here also
the kinetic terms of the mesons are generated by the quark loops. The
degrees of freedom corresponding to the mesons still have to be added
explicitly by hand. This leads to somewhat counterintuitive results
when calculating loops of mesons\rcite{BR}.
This class started as integrating the nonanomalous
variation of the measure under axial transformations and its most
recent member is known as the QCD effective action model\rcite{ERT}, that
reference also contains a rather exhaustive list of references to earlier work.

The third level differs from the previous in that it starts with
a Lagrangian which is purely fermionic and the hadronic fields are generated
by the model itself. The simplest models here are those that
add four-fermion interaction terms to the kinetic terms for the fermions.
These are usually known as extended Nambu-Jona-Lasinio (ENJL)\rcite{NJL1}
models. They have the
advantage of being very economical in the total number of parameters and of
generating the spontaneous breakdown of chiral symmetry by itself. The
previous class of models has the latter put in by hand.
Most of the remainder will be devoted to this class of models. A review
of the more traditional way of treating this model can be found in
\rcite{NJL2}.

The most ambitious method has been to find a chirally symmetric solution
to the Schwinger Dyson equations. These methods are typically plagued by
instabilities in the solution of the equations. In the end they tend to
be more or less like nonlocal ENJL models. They typically also have a lot of
free parameters. A recent reference is \rcite{Cahill}.

Some common features of all these models are that they contain a type of
constituent quark mass and confinement is introduced by hand. The quarks
are integrated out in favour of an effective action in terms of colourless
fields only. The analysis also assumes keeping only the leading term
in the expansion in the number of colours, $1/N_c$, only. This is
not always explicitly stated but there are very few papers trying to go
beyond the leading term.

I will concentrate on the ENJL models since they are the simplest ones
where the spontaneous symmetry breaking and the mesonic states are
generated dynamically rather than put in by hand. Various arguments
for this model in terms of QCD exist, see \rcite{NJL2,BBR}. A physics
argument for the pointlike fermion interaction is that in lattice calculations
the lowest glueball mass tends to be around 1~GeV. So correlations due
to gluons below this scale might be suppressed.

In the remainder I will first discuss the low-energy limit of the ENJL
model
 and then treat a method of going beyond this. The results
will be sufficiently encouraging to go on to the second part,
the use of this model to calculate hadronic matrix elements. I first give
a general introduction to the $1/N_c$ method of calculating matrix elements
and then use the ENJL model within this approach. The method will be
illustrated on the $\pi^+-\pi^0$ mass difference and on the $B_K$-parameter.

\section{Results of a Low Energy Expansion}
The basic goal is to derive the coefficients of CHPT, $f_0$, $B_0$
and the $L_i$ from an expansion in $1/m_Q^2$. $m_Q$ is the constituent quark
mass, $f_0$ and $B_0$ are related to the pion decay constant and
the quark vacuum expectation value ($\qvev$),
and the $L_i$ are the ${\cal O}(p^4)$ CHPT parameters\rcite{GL}.

The interest in this class of quark models as a low energy expansion
was restarted when in deriving the Wess-Zumino term also the effective
action produced for the ``normal'' sector was calculated. Good
agreement with the CHPT coefficients that are not connected to the
quark masses was found, see e.g. Ref.\rcite{Balog}.
This work was then extended to include the effects of ``low-energy gluons''
and quark masses\rcite{ERT}. The latter paper also used a heat kernel
method to derive the results. For the extra parameters (especially $L_5$),
the agreement was not so satisfactory.

This prompted us to examine what happens in the ENJL model\rcite{BBR}.
Its Lagrangian is a nontrivial extension of the QCD Lagrangian:
\be
\rlabel{QCD}
{\cal L}_{\rm QCD} =
\overline{q} \left\{i\gamma^\mu
\left(\partial_\mu -i v_\mu -i a_\mu \gamma_5 - i G_\mu \right) -
\left({\cal M} + s - i p \gamma_5 \right) \right\} q
 -\frac{1}{4}G_{\mu\nu}G^{\mu\nu}\ .
\ee
Here summation over colour degrees of freedom
is understood and
we have used the following short-hand notations:
$\overline{q}\equiv\left( \overline{u},\overline{d},
\overline{s}\right)$; $G_\mu$ is the gluon field in the
fundamental SU(N$_c$) (N$_c$=number
of colours) representation;
$G_{\mu\nu}$ is the gluon field strength tensor in
the adjoint SU(N$_c$) representation; $v_\mu$, $a_\mu$, $s$ and
$p$ are external vector, axial-vector, scalar and pseudoscalar
field matrix sources; ${\cal M}$ is the quark-mass matrix.

The ENJL Lagrangian is now given by the QCD Lagrangian with only
``low energy excitations'' of the quark and gluon
fields and extra pointlike interactions:
\ba
\rlabel{LENJL}
{\cal L}_{\rm QCD} &\rightarrow& {\cal L}_{\rm QCD}^{\Lambda_\chi}
+ {\cal L}_{\rm NJL}^{\rm S,P} + {\cal L}_{\rm NJL}^{\rm V,A} +
{\cal O}\left(1/\Lambda_\chi^4\right),\nonumber\\
{\rm with}\hspace*{1.5cm}
{\cal L}_{\rm NJL}^{\rm S,P}&=&
\frac{\dis 8\pi^2 G_S \left(\Lambda_\chi \right)}{\dis
N_c \Lambda_\chi^2} \, {\dis \sum_{i,j}} \left(\overline{q}^i_R
q^j_L\right) \left(\overline{q}^j_L q^i_R\right) \nonumber\\
{\rm and}\hspace*{1.5cm}
{\cal L}_{\rm NJL}^{\rm V,P}&=&
-\frac{\dis 8\pi^2 G_V\left(\Lambda_\chi\right)}{\dis
N_c \Lambda_\chi^2}\, {\dis \sum_{i,j}} \left[
\left(\overline{q}^i_L \gamma^\mu q^j_L\right)
\left(\overline{q}^j_L \gamma_\mu q^i_L\right) + \left( L \rightarrow
R \right) \right] \,.
\ea
Here $i,j$ are flavour indices and $\Psi_{R,L} \equiv
(1/2) \left(1 \pm \gamma_5\right) \Psi$.
The couplings $G_S$ and $G_V$ are
dimensionless and ${\cal O}(1)$ in the $1/N_c$ expansion and summation
over colours between brackets is understood.
The Lagrangian ${\cal L}^{\Lambda_\chi}_{\rm QCD}$ incorporates only
the low-frequency modes of quark and gluon fields. The remaining
gluon fields can be assumed to be fully
absorbed in the coefficients of the local quark field operators
or alternatively also described
by vacuum expectation values of gluonic operators (see the discussions
in Refs. \rcite{BBR,BRZ}).

So at this level we have two different pictures of this model. One is where
we have integrated out all the gluonic degrees of freedom and then
expanded the resulting effective action
in a set of {\bf local} operators keeping only the first nontrivial
terms in the expansion.
In addition to this we can make additional assumptions.
If we simply assume that these operators are produced by the short-range part
of the gluon propagator we obtain $G_S = 4 G_V = N_c\alpha_S/\pi$.
The
two extra terms in \rref{LENJL} have different anomalous dimensions,
so at the strong interaction regime, where these should be generated, there is
no reason to believe this relation to be valid. In fact the best fit
is for $G_S \approx G_V$. We report also the fit with the constraint
$G_S = 4 G_V$ included.

The other picture is the one where we only integrate out the short distance
part of the gluons and quarks. We then again expand the resulting effective
action in terms of low-energy gluons and quarks in terms of local
operators. Here we make the additional assumption that gluons only
exists as a perturbation on the quarks. The quarks feel only the interaction
with background gluons. This is worked out by only keeping the vacuum
expectation values of gluonic operators and not including propagating
gluonic interchanges. The best fits are in fact with the gluonic
expectation value equal to zero.

This model has the same symmetry structure as the QCD action
at leading order in $1/N_c$
(For explicit symmetry properties under SU(3)$_L$ $\times$
SU(3)$_R$ of the fields in this model
see reference \rcite{BBR}.) The QCD anomaly can also be consistently
reproduced\rcite{BP1}.

Numerically good agreement can be obtained for all relevant parameters
(see table \tref{table1}).
The fits are not quite as good when constraints on $G_V$ are included.
These include $G_V = 0$ and $G_V = G_S/4$. The latter follows from
requiring a kind of $SU(6)$ symmetry of the ENJL-Lagrangian and
also from the assumption that they come from one-gluon exchange. The former
is if one wants the ENJL-model to be produced by a renormalon
argument\rcite{Zakharov}.
\begin{table}
\tcaption{Best fit values for the low energy parameters using
several constraints on $G_V$.}
\begin{center}
\begin{tabular}{ccccc}
   & exp &  & & \\
\hline
$G_V$ & - &$G_S/4$& 0 & 1.264\\
$M_Q$(MeV) & - & 260 & 280 & 265 \\
$\Lambda_\chi$(MeV)& - & 810 & 630 & 1160\\
\hline
$10^3 L_2$ & 1.2 & 1.5 & 1.6 & 1.6\\
$10^3L_3$&$-$3.6 & $-$3.1 & $-$3.0 & $-$4.1 \\
$10^3L_5$ & 1.4 & 2.1 & 1.9 & 1.5\\
$10^3L_8$ & 0.9 & 0.9 & 0.8 & 0.8\\
$10^3 L_9$& 6.9 & 5.7 & 5.2 & 6.7\\
$10^3L_{10}$&$-$5.5 & $-$3.9 & $-$2.6 & $-$5.5\\
$M_V$(MeV) & 770 & 1260 &$\infty$&810\\
$M_A$(MeV)&$\approx$1260 &2010&$\infty$&1330\\
\hline
\end{tabular}
\end{center}
\label{table1}
\end{table}
The value for $G_S$ follows from the gap equation:
\be
\label{gap}
M_i = m_i - g_S \langle\overline{q}q\rangle_i
\qquad{\rm with}\qquad
\langle\overline{q}q\rangle_i = -N_c 4M_i \int\frac{d^4p}{(2\pi)^4}
\frac{i}{p^2 - M_i^2}\ ,
\ee
$m_i = 0$, $m_Q=M_i$ and $g_S=4\pi^2 G_S/(N_c\Lambda_\chi^2)$.
Typically $G_S$ is around 1.2\ .
A few examples of the formulas are:
\ba
L_9 &=& \frac{N_c}{16\pi^2}\frac{1}{6}\left[
(1-g_A^2)\Gamma(0) + 2 g_A^2 \Gamma(1)\right]\nonumber\\
f_0^2 &=& g_A \frac{N_c}{16\pi^2} 4 m_Q^2 \Gamma(0)
\ea
Here $\Gamma(0) \approx \log(\Lambda_\chi^2/m_Q^2)$ and $\Gamma(1)\approx 1$.
These functions are the result of our regularization scheme\rcite{BBR}.
The factor $g_A$ corresponds to the ``pion-quark'' axial coupling. For
$g_A\rightarrow 1$ the quark-loop model result is reproduced.
The factor $g_A$
allows for a larger $\Lambda_\chi$ for a similar $m_Q$  in the ENJL
model as compared to the chiral quark model(CQM).
In $L_9$ the last term is the CQM result while the first term corresponds
to a vector exchange. It can be seen that this model nicely interpolates
between the CQM and the vector meson dominance (VMD) result.
The largest change occurs in $L_5$ which is suppressed by an overall factor
of $g_A^3$ and a cancellation between two terms:
\be
L_5 = g_A^3\frac{N_c}{16\pi^2}\frac{1}{4}\left[\Gamma(0)-\Gamma(1)\right]\ .
\ee
This allows the ENJL result to fit the observed value for $L_5$ very well.

But much more important than the numerical results were a set of relations
that we obtained between the different low-energy parameters. These were
independent of the gluonic corrections and valid within a large class of
regularization schemes for the ENJL model.
The first set of relations is a consequence of the large $N_c$ limit:
\be
L_2 = L_1,\ L_4 = L_6 = L_7 = 0\ .
\ee
Note that at this order in $1/N_c$, the $\eta'$ is also a Goldstone boson.
The other relations are more surprising. They include the VMD relation for
the pion charge radius
\be
L_9 = \frac{1}{2}f_V g_V\ ,
\ee
and meson dominance relations for the vector and axial-vector two-point
functions
\be
L_{10}(2H_1)  = -\frac{1}{4} f_V^2 +(-) \frac{1}{4}f_A^2\ .
\ee
A similar scalar meson dominance relation holds as well:
\be
\frac{2L_5}{H_2 + 2L_8} = \frac{c_m}{c_d}\ .
\ee
For a definition of the couplings see Ref.\rcite{BBR}.
We also get a relation for the coupling $g_A$:
\be
g_A = 1 - \frac{f_\pi^2}{f_V^2 M_V^2}\ .
\ee
The most surprising result was the fact that the first Weinberg sum rule
was also among these relations:
\be
f_V^2 M_V^2 = f_\pi^2 + f_A^2 M_A^2\ .
\ee
Two more relations were violated but valid in the limit
$\Lambda_\chi\to\infty$:
\be
\label{2weinb}
f_V^2 M_V^4  =  f_A^2 M_A^4\qquad{\rm and}\qquad
M_S  = 2 M_Q\ .
\ee

\section{Beyond the Low Energy Expansion}
The success of the previous section then led us to try to go beyond
the low-energy expansion\rcite{BRZ,BP2}.
The underlying idea is not to introduce meson
like fields but to directly calculate Green functions in terms of
feynman diagrams with fermions.
Similar methods were used by Refs.\rcite{Vogl,bernard}.
As a first requisite we have to calculate the fermion propagator. This can be
done by summing the diagrams using the Schwinger Dyson equations. This
is depicted in fig. \tref{figure1}.
This leads to a constituent mass for the quarks given by eq. \rref{gap}.
\begin{figure}[htb]
\unitlength 1cm
\begin{center}
\begin{picture}(10,2)(-5,-0.5)
\thicklines
\put(-5,0){\vector(1,0){1}}
\put(-4,0){\line(1,0){1}}
\put(-2.5,-0.1){=}
\put(4,0){\line(1,0){1}}
\put(4,0.75){\circle{1.5}}
\put(4,0){\circle*{0.2}}
\thinlines
\put(3,0){\line(1,0){1}}
\put(2.3,-0.1){+}
\put(0,0){\vector(1,0){1}}
\put(1,0){\line(1,0){1}}
\end{picture}
\end{center}
\fcaption{The Schwinger Dyson equation for the propagator. A thin (thick) line
is the bare (full) fermion propagator.}
\label{figure1}
\end{figure}
In this equation we also see the close relation between
$\langle\overline{q}q\rangle$ and the constituent quark mass, $m_Q$.
To all orders we have of course a different constituent quark mass
for the different flavours. In the previous section this was treated
perturbatively.
We now will try to calculate some processes to all orders.

\subsection{Two-point functions to all orders in $q^2$ and $m_q$.}
\rlabel{twop}
The chiral limit case was analyzed in \cite{BRZ}, the corrections due to
nonzero quark masses can be found in \cite{BP2}. Several relations were
found to be true to all orders. As an example we will derive here the relation
between the scalar mass and the constituent quark mass
in the chiral limit. The set of diagrams
that contributes is drawn in Fig. \ref{Fig2pt}a. The series can be rewritten
as a geometric series and can be easily summed in terms of the one-loop
2-point function $\overline{\Pi}_S$. The full result for the scalar-scalar
two-point function (we only treat the case with equal masses here, see
\cite{BP2} for the general case) is:
\begin{equation}
\Pi_S = \frac{\overline{\Pi}_S}{1- g_S \overline{\Pi}_S}\ .
\end{equation}
The resummation has generated a pole that corresponds to a scalar particle.
Can we say more already at this level?
\begin{figure}
\begin{center}
%
%
%
\thicklines
\setlength{\unitlength}{1mm}
\begin{picture}(140.00,25.00)(-10.,15.)
\put(97.50,35.00){\oval(15.00,10.00)}
\put(08.00,33.50){$\bigotimes$}
\put(68.00,33.50){$\bigotimes$}
\put(88.00,33.50){$\bigotimes$}
\put(103.00,33.50){$\bigotimes$}
\put(17.50,35.00){\oval(15.00,10.00)}
\put(25.00,35.00){\circle*{2.00}}
\put(32.50,35.00){\oval(15.00,10.00)}
\put(40.00,35.00){\circle*{2.00}}
\put(47.50,35.00){\oval(15.00,10.00)}
\put(55.00,35.00){\circle*{2.00}}
\put(62.50,35.00){\oval(15.00,10.00)}
\put(38.50,19.00){(a)}
\put(95.50,19.00){(b)}
\put(14.50,40.00){\vector(1,0){3.00}}
\put(29.50,40.00){\vector(1,0){3.50}}
\put(44.00,40.00){\vector(1,0){5.00}}
\put(60.50,40.00){\vector(1,0){3.00}}
\put(95.50,40.00){\vector(1,0){5.00}}
\put(99.00,30.00){\vector(-1,0){3.00}}
\put(64.00,30.00){\vector(-1,0){3.00}}
\put(49.50,30.00){\vector(-1,0){3.50}}
\put(34.00,30.00){\vector(-4,1){2.00}}
\put(18.00,30.00){\vector(-1,0){2.50}}
\end{picture}
\fcaption{The graphs contributing to the two point-functions
in the large $N_c$ limit.
a) The class of all strings of constituent quark loops.
The four-fermion vertices are those of \protect{\ref{LENJL}}.
The crosses at both ends are the insertion of the external sources.
b) The one-loop case.}
\label{Fig2pt}
\end{center}
\end{figure}
We can in fact. The Ward identities for the one loop functions are:
\begin{eqnarray}
\label{WI1}
\ovpi_S &=& \ovpi_P - q^2\ovpi^{(0)}_A  \ ,\\
\label{WI2}
\ovpi_P &=& \frac{q^4}{4M_Q^2}\ovpi^{(0)}_A - \frac{\qvev}{M_Q}\ .
\end{eqnarray}
(\ref{WI1}) is a consequence of using the heat kernel for the one-loop
functions and (\ref{WI2}) is a direct consequence of the symmetry.
Using these two relations we can rewrite
\begin{equation}
1-g_S\ovpi_S = 1 + \frac{g_S}{M_Q} +
(q^2-4M_Q^2)\frac{q^2\ovpi^{(0)}_A}{4M_Q^2}
\ .
\end{equation}
The first two terms vanish due to the gap equation so this two-point function
has a pole at twice the constituent mass. For nonvanishing current
quark masses there is a small correction
\begin{equation}
M_S^2 = 4 M_Q^2 + g_A(-M_S^2) m_{ii}(-M_S^2)\ .
\end{equation}
See \cite{BP2} for definitions. Other examples of relations with the same
range of validity are:
\begin{enumerate}
\item The first and second Weinberg sum rule are satisfied, indicating a
somewhat too suppressed high energy behaviour for the last one.
\item The third Weinberg sum rule is violated as in QCD.
\item The Gell-Mann-Oakes-Renner relation to all orders in $m_q$ reads:\\
$2 m_\pi^2(-q^2) f_\pi^2(-q^2) = (m_i+m_j)(M_i+M_j)/g_S$.
\end{enumerate}
These are valid in all schemes where the one-loop functions are obtained
from a heat kernel like expansion and have thus a rather broad range of
validity. In particular, they remain valid at finite temperatures and
densities.

\subsection{3 point functions and anomalies}

We discuss in this section as an example the pseudoscalar-vector-vector
3-point function. This has all the interesting features plus the occurrence
of the flavour
anomaly. The general diagram is a one-loop triangle diagram with
a chain  with 0,1,2,3,\ldots one-loop
(like in Fig. \ref{Fig2pt}a) connected to all three corners.
The two vector legs can be easily resummed leading to
an expression like:
\begin{equation}
\frac{g_{\mu\alpha} M_V^2(-p_1^2)-p_{1\mu}p_{1\alpha}}
{M_V^2(p_1^2)-p_1^2}
\end{equation}
The resummation of the external leg leads immediately to a VMD-like
formula in terms of slowly varying functions of the momenta. A similar
expression is valid for the other vector leg.
The pseudoscalar leg can mix with the longitudinal axial-vector degree
of freedom leading to a sum of two terms. Both with a pole at
the pseudoscalar mass. Naive use of the current identity
$-iq^\alpha\ \overline{q}\gamma_\alpha\gamma_5q =
2 M_Q i \overline{q}\gamma_5q$ would lead to only the pseudoscalar term
multiplied by $g_A(-q^2)$. It is in this way that this resummation method
sees the mixing of the pion and the axial-vector.

In fact two more effects should be taken into account. The current identity
is used in a three-point function so there are usually also terms from the
equal-time commutators and there are additional terms in the current identity
due to the anomaly. These latter are very important in obtaining results that
have the correct QCD flavour anomaly \cite{BP1,BP2}.
The final result for the PVV three-point
function with momenta $p_{1,2}$ for the vector
legs is:
\begin{displaymath}
\Pi_{\mu\nu}^{PVV}(p_1,p_2) = \frac{N_c}{16\pi^2}
\frac{\varepsilon_{\mu\nu\beta\rho}p_1^\beta p_1^\rho 4 M_Q}
{g_S f_\pi^2(q^2)(m_\pi^2(q^2)-q^2)} \ \times
\end{displaymath}
\begin{equation}
\Bigg\{\frac{M_V^2(p_1^2) M_V^2(p_2^2)g_A(q^2)}{(M_V^2(p_1^2)-p_1^2)
(M_V^2(p_2^2)-p^2_2)} F(q^2,p_1^2,p_2^2)
+ 1- g_A(q^2) \Bigg\}
\end{equation}
The function $F(q^2,p_1^2,p_2^2)$ is essentially the chiral quark loop result.
So analytically we only have a part that is multiplied by the expected
Vector-Meson-Dominance factors. There is a second part that is not, that
came from the extra terms in the current identity. This behaviour is in fact
very welcome. We have both the one-loop quark contribution to the slopes
and the one from vector meson dominance. Since both of these explain the
observed slopes having both fully would not agree with experiment.
Here we have, however,
$F_{PVV}(m_\pi^1,p_1^2,p_2^2)$
$\approx$ $1 +$ $\rho(p_1^2+p_2^2)$ $+\rho^\prime m_\pi^2 +\cdots$
with $\rho
=g_A(0)\left(1/M_V^2(0)+ 1/(12M_Q^2)\right)$$\approx 1.53~{\rm GeV}^{-2}$
and
$\rho^\prime =$ $(g_A(0)/(12M_Q^2))\left(1-\Gamma(1)/\Gamma(0))
\right)$ $\approx$ $0.40$ ${\rm GeV}^{-2}$.
As we see we have good numerical agreement with the observed slope and the
corrections due to finite meson mass are substantially smaller than in the
chiral quark model. The latter is also desirable since otherwise there would
have been extremely large corrections to the $\eta$ decay.

\subsection{Meson Dominance}

As shown in the previous two subsections the appearance of meson dominance
like formulas with slowly varying couplings is a natural feature
of this model and as such the successful phenomenology of this concept
is taken over. The model does combine this together with a set of
chiral quark loop effects in a kind of interpolating fashion thus
incorporating the strengths of both approaches.
The final results can be plotted to check whether the final formulas
also have a VMD-like behaviour and as shown in section 5 in \cite{BP2}
this is numerically the case for all the 2 and 3-point functions
studied there.
We have in general stayed in the euclidean domain of momenta to avoid the
problem that this model does not include confinement. There the sign of
meson dominance is that inverse formfactors are straight lines as a function
of the various $q^2$. This we find indeed.

\section{The $\pi^+-\pi^0$ Mass Difference}

In this section we will discuss the general philosophy behind  the $1/N_c$
method of calculating nonleptonic matrix elements. A good review where
also the references to the original papers can be found are the lecturers
by G\'erard\rcite{Gerard}. The application of this method to the
$\pi^+-\pi^0$ mass difference can be found in Ref.\rcite{BBG}
and the calculation within the QCD effective action model and the ENJL
model is in Refs.\rcite{BR,BRZ}.

We look at this quantity because it is the simplest nonleptonic
matrix elements is several respects. There is no factorizable
contribution because the photon is spin 1 and the pion spin 0. It
involves only pions so we expect the limit where the current quark masses
vanish to be a good approximation and (unlike $B_K$) it doesn't vanish and is
well defined in this limit. The latter remark has one very useful consequence.
Using PCAC it can be shown\rcite{Das} that this matrix element can be related
to a vacuum matrix element. So the mass difference becomes a
vacuum matrix element
of the photon propagator integrated over all momenta in the presence
of the strong interactions.
Schematically, the matrix element $\langle\pi^+|J^2|\pi^+\rangle$
can be rewritten in terms of
$\langle 0|J^2|0\rangle$. The precise expression in terms of
the hadronic two-point functions is given by\rcite{Das,BBG,BR}
\be
\label{dass}
m_{\pi^+}^2 - m_{\pi^0}^2 =
-\frac{3\alpha_{em}}{8\pi f_\pi^2}
\int^\infty_0 dQ^2\ Q^2\left(
\Pi^{(1)}_V(Q^2)-\Pi^{(1)}_A(Q^2)\right)\ .
\ee
Eq. \rref{dass}
involves an integral over all distance scales. The underlying idea
is now to split this integral into two parts,
$
\int_0^\infty = \int_0^{\mu^2} + \int_{\mu^2}^\infty
$,
and then to evaluate both pieces separately.

The long distance part in $1/N_c$ can be calculated in models since in $1/N_c$
the only quantities needed are the couplings of currents to hadrons and not
of full four-quark operators to hadrons.
The essence of the $1/N_c$ method is to do the short-distance part using the
operator expansion and then use $1/N_c$ to evaluate the matrix element.
Here this corresponds to using as the
difference of 2-point
functions:
\be
\left(
\Pi^{(1)}_V(Q^2)-\Pi^{(1)}_A(Q^2)\right)
=
\frac{-1}{Q^6}{ 8\pi \alpha_S(Q^2)} \left(\qvev\right)^2\ .
\ee
This leads to\rcite{BBG}
\be
\left.\Delta m_\pi^2\right|_{\rm SD} = \frac{3\alpha\alpha_S}{f_\pi^2 \mu^2}
\left(\qvev\right)^2\ .
\ee
One can then still do a renormalization group improvement of this\rcite{BRZ}.

\begin{figure}
\hspace{1.5cm}\epsfxsize=12cm\epsfbox{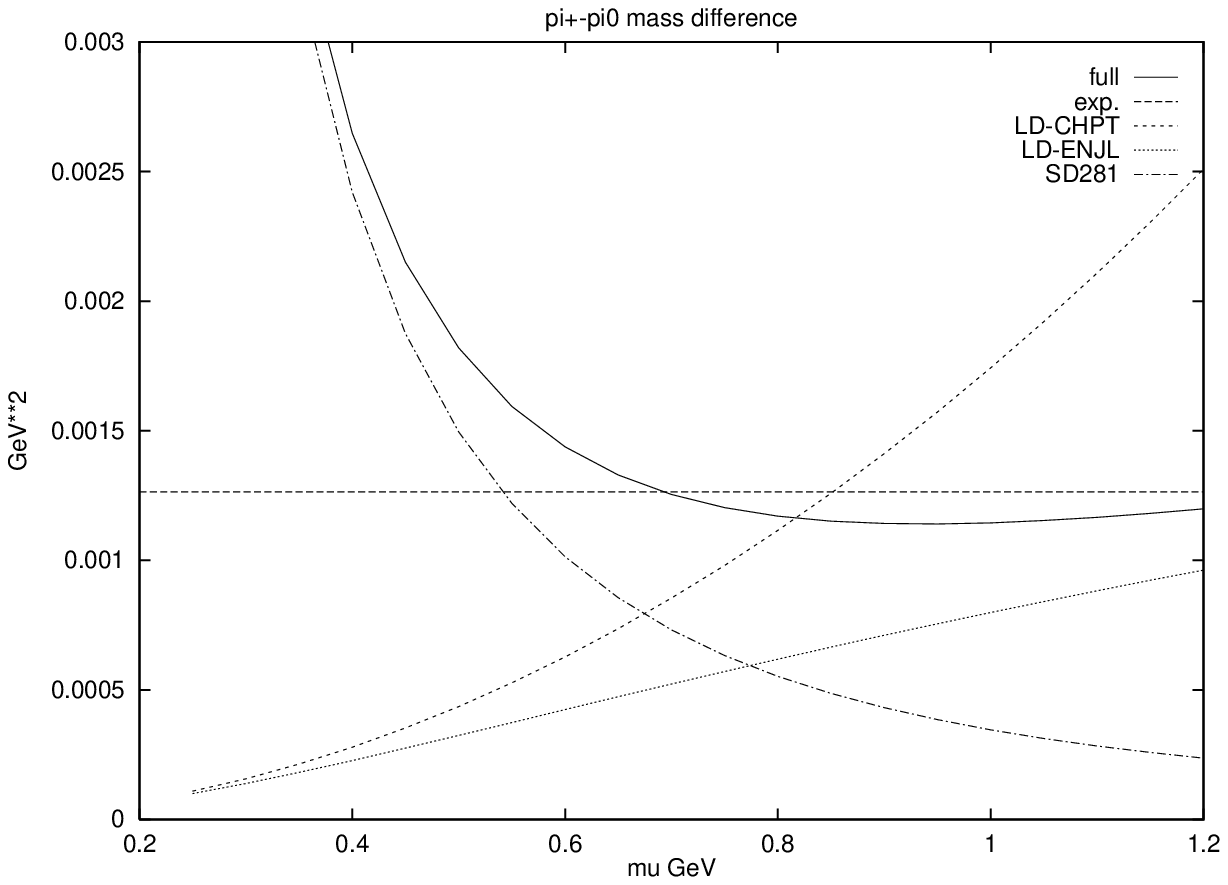}
\fcaption{The results for $m_{\pi^+}^2-m_{\pi^0}^2$:
long-distance result with only the pion (LD-CHPT);
ENJL long-distance (LD-ENJL); experimental
value (exp.); the short-distance result (SD281)
and the sum of short- and long-distance ENJL (full).}
\label{figure3}
\end{figure}

The long distance part of the integral requires more care. There are several
approaches.
\begin{enumerate}
\item One can take the measured spectral functions and use these to evaluate
the two-point functions needed in the integral. The most recent evaluation
of this is in Ref.\rcite{Don2}.
\item The two-point functions can be approximated by including the
$\rho$, $\pi$ and $a_1$ contribution. This was done neglecting the QCD part
in the original paper\rcite{Das} and more recently in \rcite{BBG}.
\item We can take only the $\pi$ contribution\rcite{BBG}.
This is most like the original
$1/N_c$ method(Ref.\rcite{Gerard} and references therein). This leads to
$\Delta m_\pi^2 = 3\alpha\mu^2/(4\pi)$.
\item One can use the QCD effective action approach\rcite{BR}.
\item The ENJL model can be used\rcite{BRZ}.
\end{enumerate}
All of these approaches give a good result for the mass difference.
In cases 1,2 and 5 a good matching was also obtained. This means that we
can vary $\mu$, the split between the short- and long-distance part of the
integral, over a reasonable interval without changing the result. In
figure \tref{figure3} the long-distance result with only the pion
is shown and the ENJL long-distance result. Also shown is the experimental
value, the short-distance result and the sum of short- and long-distance
for the ENJL case. The value of $\qvev$ used is the one given by the
ENJL model.

At this point I would like to remark that for this quantity in the QCD
effective
action approach one only obtains a gauge invariant result if the pion
is explicitly taken as propagating (see\rcite{BR}). This shows that in
this model the pion degree of freedom has to be added by hand. The gauge
dependence then cancels between a two- and a three-loop diagram.

\section{$B_K$}
In this section the extension to weak nonleptonic matrix elements of
the methods in the previous section is discussed on the example
of $B_K$. Here again the pure $1/N_c$ method\rcite{Gerard,BBG2},
the QCD effective action model\rcite{AP} and the ENJL model\rcite{BP3}.
An overview of theoretical situation a few years ago can be found in
Ref.\rcite{Buras1}. The main alternatives to the present method are lattice
calculations\rcite{lattice} and 2 and 3-point QCD sum rules\rcite{QCD1}.

The short-distance integration here is done using the renormalization group.
This sums the possible large logarithms involving $\log(m_W^2/\mu^2)$.
The problem then reduces to the study of
\be
\rlabel{defbk}
\langle \overline K^0  | {\cal O}_{\Delta S=2} (x)
| K^0  \rangle \equiv \frac{\dis 4}{\dis 3}
B_K (\mu) f^2_K m_K^2
\ee
with
the $\Delta S=2$ operator
$
{\cal O}_{\Delta S=2} (x)  \equiv L^{sd}_\mu(x) L_{sd}^\mu (x)
$;
$2L^{sd}_\mu(x) = \overline s (x) \gamma_\mu
\left( 1-\gamma_5\right) d (x)$ and
summation over colours is understood.
Eq. \rref{defbk} is also the definition of the $B_K$ parameter.
The different approximations give
\begin{enumerate}
\item Vacuum Insertion : $B_K(\mu) = 1$.
\item Leading in $1/N_c$: $B_K(\mu) = 3/4$.
\item The standard $1/N_c$ result: $B_K(\mu) = 3/4(1-2\mu^2/(16\pi^2f_\pi^2))$.
\item $1/N_c$ with inclusion of vector mesons\rcite{Gerard}:
\be
 B_K(\mu) = \frac{3}{4}\left(1-\frac{1}{16\pi^2f_\pi^2}
\left(
-\frac{7}{8}\mu^2-\frac{3}{4}\frac{m_V^2\mu^2}{\mu^2+m_V^2}
-\frac{3}{8}m_V^2\log\frac{\mu^2+m_V^2}{m_V^2}\right)\right)\ .
\ee
\item The QCD effective action result\rcite{AP}:
\be
B_K(\mu)=\frac{3}{4}\left(1+\frac{1}{N_c}\left(
1-\frac{N_c}{32\pi^2 f_\pi^4}\langle\frac{\alpha_S}{\pi}G^2\rangle+\ldots
\right)\right)\ .
\ee
\end{enumerate}
We would also like to study the effects of off-shellness. Therefore we do not
directly study the matrix element in Eq. \rref{defbk} but the Green function
\be
\rlabel{twopoint}
G_F\,\Pids(q^2)  \equiv
 i^2 \int d^4 x \, e^{iq\cdot x}
\langle 0 | T \left( P^{ds}(0)P^{ds}(x) \Gamma_{\Delta S =2}
\right)| 0 \rangle
\ee
in the presence of strong interactions. We use the ENJL model
for scales
below or around the spontaneous symmetry breaking scale. Here
$G_F$ is the Fermi coupling constant, we use
$P^{ds}(x) = \overline d (x)i\gamma_5 s (x)$, with summation over colour
understood and
$
\rlabel{operator}
\Gamma_{\Delta S=2} = - G_F\,
\int d^4 y \, {\cal O}_{\Delta S = 2} (y)
$.
 The reason to calculate this two-point function rather than
directly the matrix element is that we can now perform the calculation
fully in the Euclidean region so we do not have the problem
of imaginary scalar products. This also allows us in principle to
obtain an estimate of off-shell effects in the matrix elements. This
will be important in later work to assess the uncertainty when trying
to extrapolate from $K\to\pi$ decays to $K\to 2\pi$.
This quantity is also very similar to what is used in
the lattice and QCD sum rule calculations of $B_K$.

The $\Delta S = 2$ operator can be
rewritten as
\be
\rlabel{operator3}
\Gamma_{\Delta S=2} = - G_F \, \int \frac{d^4 r}{(2\pi)^4}
\int d^4 x_1 \int d^4 x_2 \,
e^{-i r \cdot(x_2 - x_1)} L^{sd}_\mu(x_1) L_{sd}^\mu(x_2) .
\ee
This allows us to consider this operator as
being produced at the $M_W$ scale
by the exchange of a heavy $X$ $\Delta S = 2$ boson.
We will work in the Euclidean domain
where all momenta squared are negative.
The integral in the modulus of the momentum $r$ in
\rref{operator3} is then split into two parts,
$
\int_0^{M_W} d |r| = \int_0^\mu d |r| + \int_\mu^{M_W} d |r|
$.
In principle one should then evaluate both parts separately as was done
for the $\pi^+-\pi^0$ mass difference
in the above quoted references. Here we will do
the upper part of the integral using the renormalization group.
This results in the integral being of the same form but multiplied
with the Wilson coefficient $C(\mu)$,
\be
\rlabel{operator2}
\Gamma_{\Delta S=2} = - G_F \, C(\mu) \int_0^\mu \frac{d^4 r}{(2\pi)^4}
\int d^4 x_1 \int d^4 x_2 \, e^{-ir\cdot(x_2 - x_1)}
L^{sd}_\mu(x_1) L_{sd}^\mu (x_2)\ .
\ee

This can now be studied using the $1/N_c$ expansion.We can first do this within
a chiral expansion leading to the result
in the chiral limit (see Ref.\rcite{BP3} for details):
\be
\rlabel{bknlo}
B_K(\mu)_{CHPT} =
\frac{3}{4}\left( 1 -\frac{1}{16\pi^2 f_0^2}
\left[2 \mu^2 + \frac{\dis q^2}{\dis 2}\right] \right)\ .
\ee
The correction is negative. It disagrees somewhat with the result obtained
in \rcite{BBG2}
because there no attempt at identifying the cut-off
across different diagrams was made.
Since we work at leading level in $1/N_c$
in the NLO CHPT corrections we have included the relevant
singlet ($\eta_1$) component as well using nonet symmetry.
The correction in \rref{bknlo} has precisely
the right behaviour to cancel partly $C(\mu)$ which increases with
increasing $\mu$.

The same calculation can now be performed for the ENJL model. Here the major
complication is the number of different diagrams that has to be evaluated.
An example of one of the classes is shown in figure \tref{figure4}.
\begin{figure}[htb]
\hspace{3.5cm}\epsfxsize=8cm\epsfbox{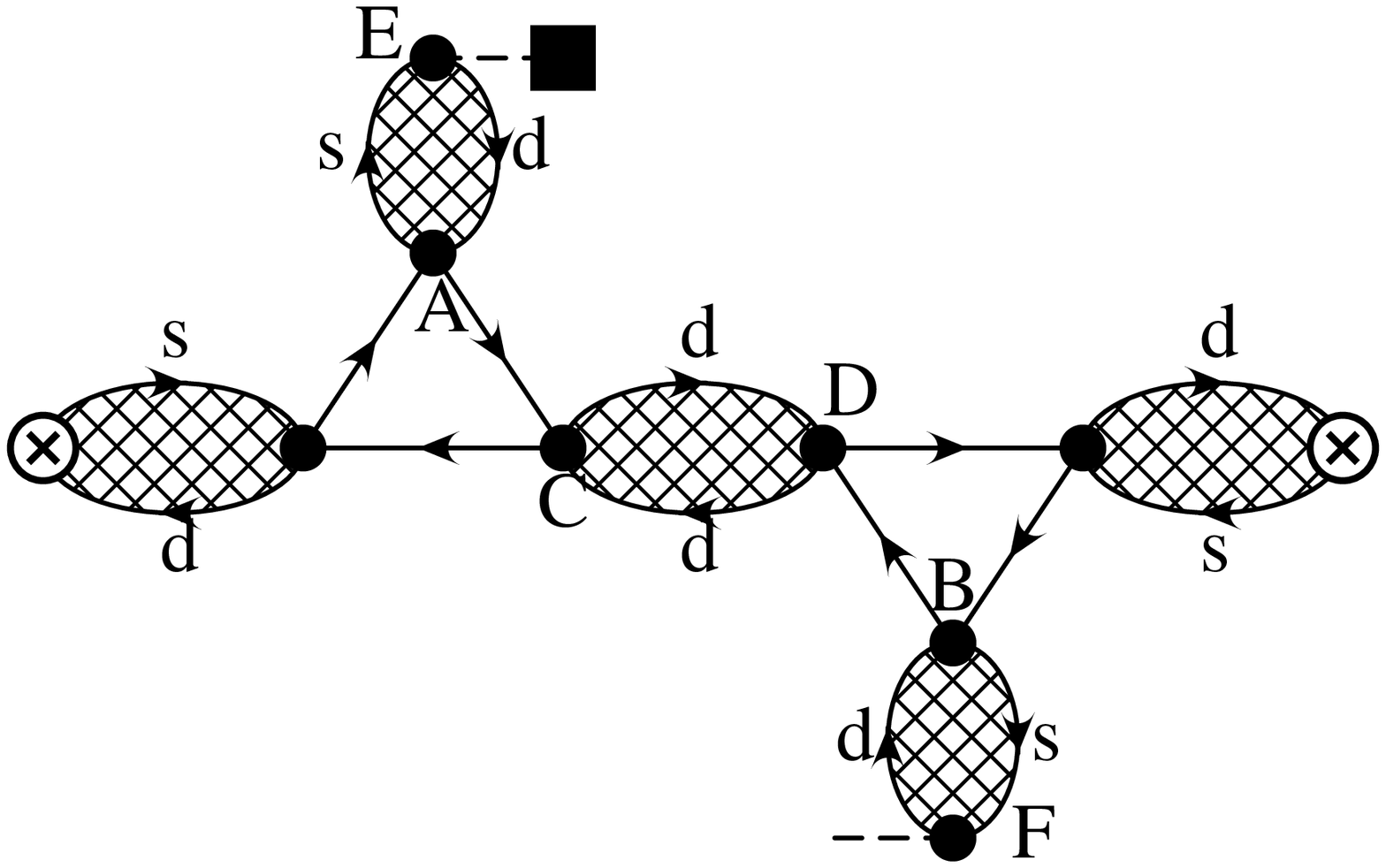}
\fcaption{A leading $1/N_c$ contribution to
the nonfactorizable part of $\Pids(q^2)$
in the NJL model. The crosshatched areas are the full two-point functions
as discussed in subsection 3.1. Point E and F are connected via
$\Gamma_{\Delta S=2}$.}
\rlabel{figure4}
\end{figure}

We now evaluate all contributions numerically to the two-point function
of Eq. \rref{twopoint}. The results for several input values are
in table \tref{table2}.
\begin{table}[hbt]
\tcaption{Results for $B_K$ and $\hat B_K$ in the ENJL model.}
\begin{center}
\begin{tabular}{c|cc|ccc|cc}
$\mu$ (GeV) & $B_K^\chi(\mu)$ & $\hat B{}_K^\chi$ & $B_K^m(\mu)$ &
$B_K^a(\mu)$ &
 $\hat B_K^m$ & $B_K^{\rm eq}(\mu)$ & $\hat B_K^{\rm eq}$ \\
\hline
0.3  & 0.68 &0.50 & 0.74&   0.50&0.55  &0.74 &0.55 \\
0.5  & 0.59 &0.59 & 0.71&$-$0.44&0.71  &0.72 &0.72 \\
0.7  & 0.53 &0.58 & 0.69&$-$2   &0.75  &0.68 &0.75 \\
0.9  & 0.48 &0.55 & 0.66&$-$3   &0.76  &0.65 &0.75 \\
1.1  & 0.45 &0.54 & 0.64&$-$4   &0.76  &0.64 &0.76
\end{tabular}
\end{center}
\rlabel{table2}
\end{table}
We have studied three cases, namely, the chiral case, $m_d = m_s =0$,
the case with SU(3) symmetry breaking $m_s = 83 ~{\rm MeV} \ne m_d
= 3~{\rm MeV}$  and the case with $m_s = m_d = 43~{\rm MeV}$.
The other parameters are
$G_S = 1.216$, $\Lambda_\chi = 1.16~{\rm GeV}$ and $G_V = 0$.
The latter simplifies the calculation by about an order of magnitude.
Preliminary results for the $G_V\ne 0$ case have the same qualitative
conclusion\rcite{BP4} but typically somewhat lower values of $B_K$ and
less good matching.

The procedure we have followed to analyze the numerical results
is the following.  We fit the ratio between the correction and the
leading $1/N_c$
result for a fixed scale $\mu$ to $a/q^2 + b + c q^2$ which always gives
a very good fit ($a$, $b$ and $c$ are $\mu$ dependent).
 Once we have this fit we can extrapolate our
$B_K$ form factor (remember that we have calculated it for
Euclidean $q^2$) to the physical $B_K$, i.e. to
$q^2=m_K^2\approx 0,0.13~{\rm GeV}^2$ (chiral,other cases).

Let us first treat the chiral or massless quarks case.
Here a nontrivial check on the
results is that the diagrams have a behaviour which sums to $1/q^2$,
i.e. $a$ should be zero.
The individual contributions do not have this behaviour.
$b$
is the relevant contribution to $B_K$ since $m_K^2|_\chi=0$.
The first three
columns in table \tref{table2} are $\mu$, $B_K^\chi(\mu)$ and
$\hat B_K^\chi = B_K(\mu)\alpha_S(\mu)^{a_+}$ with
$a_+ = -2/9$ and $\Lambda^{(3)}_{\overline{MS}} = 250$ MeV. The hatted
quantity is the scale independent quantity. Good matching is obtained if
this value is stable within a range of $\mu$.

In the second case ($m_s \ne m_d$),
because the chiral symmetry is broken, there is
a possibility for contributions to $B_K(q^2)$ that are not
proportional to $q^2$, i.e. $a\ne 0$.
In fact a CHPT calculation predicts
precisely the presence of this type of terms\rcite{BP4}. For
small values of $q^2$ the part due to $a$ dominates even though it is only a
small correction when extrapolating to the physical $B_K^m$ at $q^2=m_K^2$.
This can be found in column 4.
The fifth column is the form factor $B_K^a$ for $q^2=-0.001$
 GeV$^2$ where the correction due to the $a$ term is sizeable.
Notice the difference
between these two columns.
This same feature should be visible in the lattice
calculations as soon as they are done with different
quark masses. The invariant $\hat B_K^m$ for this case is in column 6.

In the the last case, i.e. $m_d=m_s$, which is similar to the present
lattice QCD calculations,  the fit gives $a=0$ to a good
precision and  the value of $B_K^{\rm eq}$ extracted is rather
independent of $q^2$. The invariant $\hat B_K^{\rm eq}$ in this case is in
column 8.

In view of the results of \rcite{BBR,NJL2,BP2} we expect to get
a good prediction for the effects of non-zero and different current quark
masses.
We see those and find a significant
change due to both:
\be
\hat B_K^m(m_K^2\approx 0.13~{\rm GeV}^2) \approx 1.35
 \, \hat B_K^\chi(m_K^2= 0)
\ee
for scales $\mu\approx (0.7\sim 1.1) ~{\rm GeV}$.
For the extrapolation to the kaon pole the
difference between the masses has a much smaller effect than the fact that
they were non-zero. In order to compute $B_K$ in the general case a careful
extrapolation to the poles was needed. The final correction to the $B_K$
parameter compared to its leading value of $3/4$ turns out to be rather
small.

\section{Conclusions}
To leading order in $1/N_c$ hadronic parameters can be reasonably well
understood from constituent quark models. A good choice is the ENJL model,
it combines few parameters with good predictions and a generation of
the meson fields dynamically. The use of this model for nonleptonic
matrix elements has given first results for $B_K$ and $\Delta m_\pi^2$.

\section{Acknowledgements}
I would like to thank the organizers for a most enjoyable meeting and stay
in Croatia.

\end{document}